\documentclass[article,12pt,onecolumn,showpacs,preprintnumbers,amsmath,assume,plain]{revtex4}
\usepackage{graphicx}
\usepackage{dcolumn}
\usepackage{bm}
\usepackage{hyperref}

\begin{document}

\title{Inversion of walkaway VSP data in the presence of \\ lateral velocity heterogeneity}

\author{Vladimir Grechka$^{1,3}$, Ilya Tsvankin$^{1}$, and Pedro Contreras$^{2}$}
\affiliation{$^{1}$Center for Wave Phenomena, Department of
Geophysics,Colorado School of Mines, Golden, CO 80401}
\affiliation{$^{2}$PDVSA-INTEVEP, Apartado 76343, Caracas
1070A, Venezuela}
\affiliation{$^{3}$presently at Marathon Oil}
\date{\today}

~\vspace{1cm}

\begin{abstract}
{Multi-azimuth walkaway vertical seismic profiling (VSP)
is an established technique for the estimation of in situ slowness
surfaces and inferring anisotropy parameters.} Normally, this
technique requires the assumption of lateral homogeneity, which
makes the horizontal slowness components at depths of downhole
receivers equal to those measured at the surface. Any
violations of this assumption, such as lateral heterogeneity or
nonzero dip of intermediate interfaces, lead to distortions in
reconstructed slowness surfaces and, consequently, to errors in
estimated anisotropic parameters.

Here, we relax the assumption of lateral homogeneity and
discuss how to correct VSP data for {\em weak}$\,$ lateral
heterogeneity (LH). {We describe a procedure of downward
continuation of recorded traveltimes that accounts for the
presence of both vertical inhomogeneity and weak lateral
heterogeneity}, which produces correct slowness surfaces at
depths of downhole receivers.
Once the slowness surfaces are found and a desired type of
anisotropic model to be inverted is selected, the corresponding
anisotropic parameters, providing the best fit to the estimated
slownesses, can be obtained. We invert the slowness surfaces of
$P$-waves for parameters of the simplest anisotropic model
describing dipping fractures -- transversely isotropic medium
with a tilted symmetry axis. {Five parameters of this model --
the $P$-wave velocity $V_0$ in the direction of the symmetry
axis, Thomsen's anisotropic coefficients $\epsilon$ and
$\delta$, the tilt $\nu$, and the azimuth $\beta$ of the
symmetry axis can be estimated in a stable manner when maximum
source offset is greater than half of receiver depth.}

\end{abstract}

 \maketitle

\section{Introduction}

Obtaining anisotropic velocity fields is one of the main
challenges in extending seismic processing to anisotropic
media. Although analysis of $P$-wave surface reflection data
allows one to estimate subsets of anisotropic parameters
sufficient for time processing, the vertical velocity usually
remains undetermined in such important for exploration
anisotropic models as transversely isotropic with a vertical
symmetry axis (VTI) \protect\cite{alk95} and orthorhombic (ORT)
\protect\cite{gre97,con2}. Lack of information about vertical
velocity leads to distortions in a vertical scale of
depth-migrated seismic sections, routinely
observed in areas with non-negligible anisotropy. One possible
way to obtain the true vertical velocity and, therefore, correct
depth images in VTI and ORT media
\protect\cite{alk95,con3} is to explicitly measure velocity
using check shots or near-offset vertical seismic profiling
(VSP) data.

\begin{figure}[htp!]
\includegraphics[width = 5.0 in]{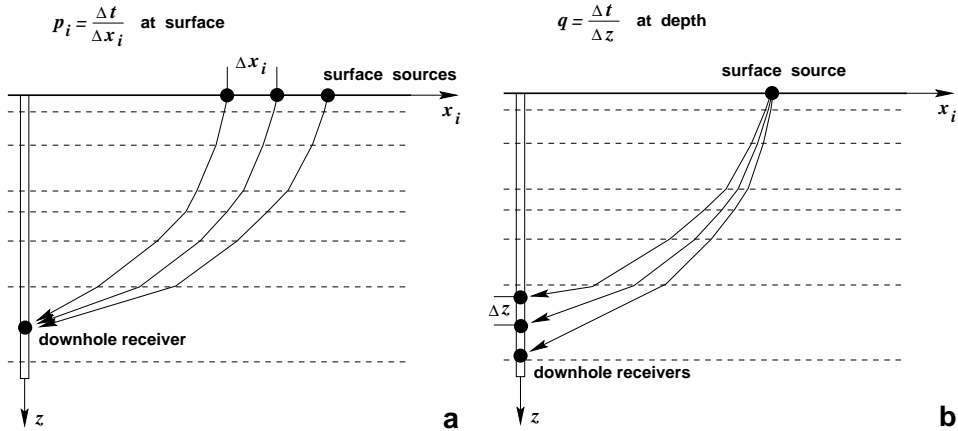}
\caption{Principal scheme of calculating (a) the horizontal
components $p_i$ ($i=1,2$) of the slowness vector and (b) its vertical component $q$ \protect\cite{gai90}.}
\label{fig01}
\end{figure}

{Traveltimes of direct $P$-arrivals recorded in
multi-azimuth walkaway VSP geometry provide information not
only about true vertical velocity but also about some portion
of the slowness surface, which can be used to reconstruct in
situ anisotropic parameters}. The principal scheme of obtaining
the slowness surface \protect\cite{gai90} is shown in
Figure~\ref{fig01}. The horizontal $p_i$ ($i=1,2$) and vertical
$q$ components of the slowness vector are, by definition, the
components of traveltime gradient, i.e.,
\begin{equation}
   p_i = \frac{\partial t}{\partial x_i} \qquad {\rm and} \qquad
   q = \frac{\partial t}{\partial z} \, .
\label{eq01}
\end{equation}

Therefore, having measured traveltimes between several surface
sources located along coordinate axes $x_1$ and $x_2$ of a
selected coordinate frame and downhole receiver at the depth
$z$, we can calculate the horizontal slowness components $p_i$
(Figure~\ref{fig01}a). Using traveltimes recorded for a given
source and several downhole receivers, we can find the vertical
slowness component $q$ (Figure~\ref{fig01}b). The problem,
however, is that the horizontal slowness components are
measured {\em at the surface}$\,$ while the vertical component
is obtained {\em at the receiver depth}. If medium above
receiver is laterally homogeneous, the measured horizontal
slowness components $p_i$ are preserved along downward
propagating rays due to Snell's law and become equal to those
at receiver depth, thus, giving local value of slowness $q(p_1,
p_2)$ at the depth $z$. The assumption of lateral homogeneity
is conventionally made in the papers on estimation of
anisotropic parameters from VSP data
\protect\cite{gai90,mil94,mil294}.

On the other hand, it is known that if this assumption is
violated due to the presence of lateral heterogeneity or
nonzero dip of intermediate interfaces, erroneous values of $q$
as function of $p_1$ and $p_2$ are obtained
\protect\cite{gai90}. For example, Sayers in 1997 \protect\cite{say97} found
that dip of an intermediate interface of only $5^\circ$
significantly distorts $q(p_1, p_2)$ values.

Here, we show that information about LH contained in
traveltimes recorded in VSP geometry can be, under certain
circumstances, extracted and the influence of LH on traveltimes
can be removed. The assumptions we make are the following:
\begin{enumerate}
	\item The medium is vertically homogeneous at each interval
	between downhole receivers.
	\item Anisotropy is factorized
	with respect to lateral coordinates, i.e., anisotropy
	coefficients are constant at each interval while velocity
	itself may vary laterally.
	\item Lateral heterogeneity is weak.
\end{enumerate}

{The first two assumptions allow us to separate the
influence of LH on traveltimes from those of vertical
inhomogeneity and anisotropy; the third assumption makes it
possible to linearize traveltimes with respect to LH and derive
explicit equations expressing the contribution of LH.} Based on
these equations, we develop a procedure to propagate the recorded
traveltimes downward and compute slowness surfaces $q(p_1,
p_2)$ at receiver depths. We present numerical examples
illustrating improvements in accuracy of reconstructed slowness
surfaces compared to those obtained using the conventional
approach.

To invert anisotropy parameters from the best fit of computed and
extracted slowness surfaces, a certain anisotropic model has to
be chosen. Usually, only $P$-wave slownesses are obtained and
inversion is performed for models with horizontal symmetry
plane -- VTI \protect\cite{gai90} or orthorhombic
\protect\cite{mil94}. Since this sort of inversion does not
require anisotropic medium to posses a horizontal symmetry
plane, we assume the model to be transversely isotropic with a
{\em tilted}$\,$ symmetry axis (TTI model) and estimate its
five parameters: the $P$-wave velocity $V_0$ in the direction of
the symmetry axis, generic \protect\cite{thom} anisotropic
coefficients $\epsilon$ and $\delta$, and the tilt $\nu$ and
azimuth $\beta$ of the symmetry axis.

\begin{figure}[ht]
\includegraphics[width = 4.0 in]{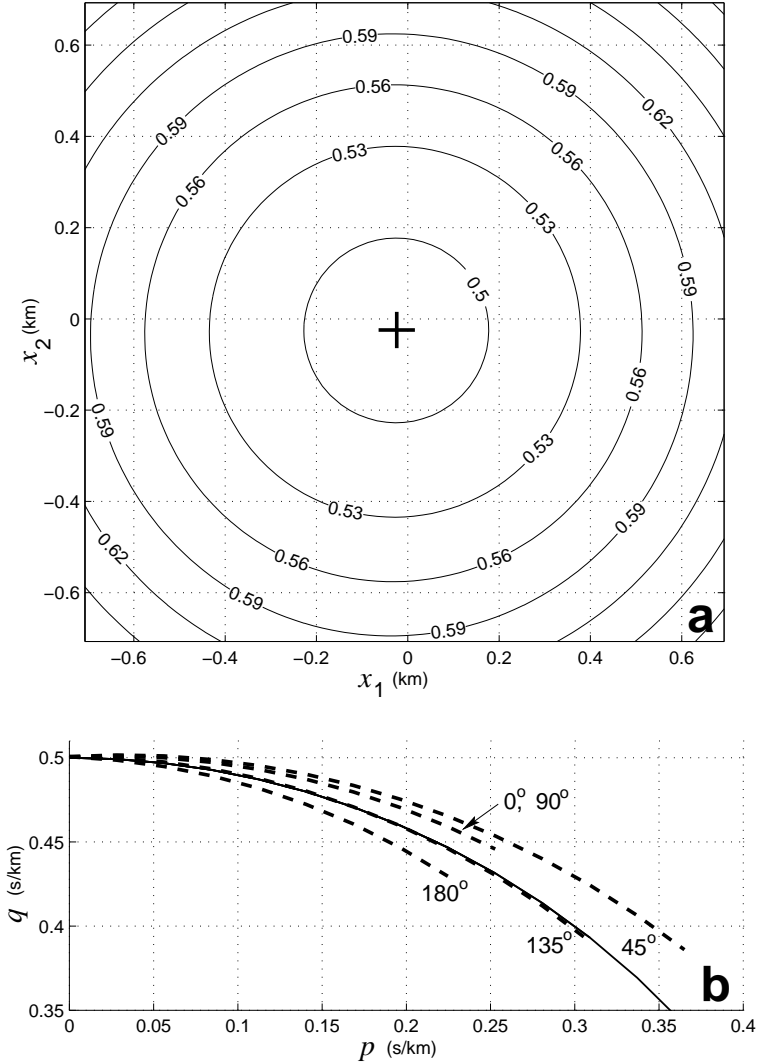}
\caption{(a) Contours (in s) of traveltime surface $t(x_1, x_2)$ recorded at
         depth $z = 1$~km in isotropic model with laterally varying velocity
         $V_0(x_1, x_2) = 2.0 - 0.1 (x_1 + x_2)$ [km/s].
         The sign ``{\bf +}'' indicates the position of traveltime
         minimum.
         (b) Cross-sections of the slowness surface $q(p_1, p_2)$, where
         $p_1 = p \cos \alpha$ and $p_2 = p \sin \alpha$, at azimuths
         $\alpha = 0^\circ, \, 45^\circ, \, 90^\circ, \, 135^\circ$, and
         $180^\circ$ (with respect to axis $x_1$) reconstructed under the
         assumption of lateral homogeneity (dashed) and
         cross-section of the correct isotropic slowness surface (solid).
        }
\label{fig02}
\end{figure}

\section{Ambiguity between anisotropy and lateral heterogeneity}
We begin our discussion with an example illustrating that
lateral heterogeneity, even perceived as insignificant, does
present a problem for conventional approach of estimating
slowness surfaces and lead to erroneous conclusions about
anisotropy of the subsurface. Figure~\ref{fig02}a shows the
contours of $P$-wave traveltime surface $t(x_1, x_2)$ recorded
by the receiver at depth $z = 1$~km located in vertical
borehole with coordinates $[x_1, x_2] = [0, 0]$~km. The
traveltimes were computed numerically using the technique
described by Grechka and McMechan \protect\cite{gre3} in a purely
isotropic model with linear lateral velocity variation
$V_0(x_1, x_2) = 2.0 - 0.1 (x_1 + x_2)$ [in km/s]. Velocity
heterogeneity in this model can be characterized by the
absolute value of lateral velocity gradient $h = 0.1$~s$^{-1}$
or by the velocity variance which is equal to only 2.7\% for
the offsets $x_1$ and $x_2$ shown in Figure~\ref{fig02}a.
{Nevertheless, lateral heterogeneity clearly manifests
itself by shifting the traveltime minimum (marked with the plus
in Figure~\ref{fig02}a) away from the zero offset. In
principle, this shift can be attributed to any combination of
the following three factors: 1) lateral heterogeneity in the
subsurface; 2) anisotropy without horizontal symmetry plane;
and 3) uncertainty in the lateral receiver position. We assume the
receiver location to be known exactly and analyze the
influences of the first two factors.}

Conventionally, one would assume lateral homogeneity and
explain the shift of the traveltime minimum from the zero offset by
anisotropy. Following this approach, we calculated the
derivatives $p_i = \partial t / \partial x_i$ of the traveltime
shown in Figure~\ref{fig02} and estimated $q = \partial t /
\partial z$ using traveltimes computed for the array of
receivers at depths $z = [0.98, 0.99, 1.00, 1.01, 1.02]$~km.
Then, we reconstructed the slowness surface $q(p_1, p_2)$. Its
cross-sections along several azimuthal directions, shown in
Figure~\ref{fig02}b, clearly indicate the presence of apparent
{\em azimuthal anisotropy}. Note that the deviation from the
correct slowness surface (solid line in Figure~\ref{fig02}b) is
the greatest in the direction of lateral velocity gradient
(azimuth $\alpha = 45^\circ$) and smallest in the orthogonal
direction at azimuth $\alpha = 135^\circ$. This observation is
analogous to the result of Sayers (1997) \protect\cite{say97},
who studied the influence of dipping interfaces in the
overburden on recovering the slownesses in VTI media and
concluded that the influence of dip is negligible for
acquisition in the strike direction
The next question, which might be asked, is whether one can find an
anisotropic model that explains the cross-sections of $q(p_1,
p_2)$ surface in Figure~\ref{fig02}b. This model has to be
azimuthally anisotropic as obvious from Figure~\ref{fig02}b.
Also, it is not supposed to have a horizontal symmetry plane
because otherwise it would be impossible to explain the shift
of the traveltime minimum seen in Figure~\ref{fig02}a. The
simplest anisotropic model satisfying both requirements is
TTI -- transversely isotropic with a tilted symmetry
axis. The $P$-wave slowness surface in this model, being
insensitive to the shear-wave velocity (Grechka and Tsvankin, 1998)
\protect\cite{gre98}, is determined by five quantities: the
$P$-wave velocity $V_0$ in the direction of the symmetry axis,
Thomsen's (1986) coefficients $\epsilon$ and $\delta$, the tilt
$\nu$, and the azimuth $\beta$ of the symmetry axis. We invert
the vector $\chi \equiv [V_0, \epsilon, \delta, \nu, \beta]$
containing those five quantities by minimizing the
least-squares objective function
\begin{equation}
   {\cal F}(\chi) =
   \left[ \frac
          {\sum_{j=1}^N \left( q(p_1^{(j)}, p_2^{(j)}) -
                              \tilde{q}(p_1^{(j)}, p_2^{(j)}, \chi)
                       \right)^2 }{N-1}
   \right]^{\frac{1}{2}} \,
   \label{eq02}
\end{equation}
that expresses the standard deviation of $N$ computed vertical
slownesses $\tilde{q}$ from the vertical slownesses $q$
measured from VSP traveltimes. The slownesses
$\tilde{q}(p_1^{(j)}, p_2^{(j)}, \chi)$ are calculated from the
Christoffel equation \protect\cite{mus}. The minimization of
the objective function ${\cal F}$ is carried out using the
simplex method (Press et al., 1987) \protect\cite{nr}.

Table~1 shows parameters of the obtained TTI model along with
parameters of the true isotropic model. Although the values of
velocity $V_0$ in the direction of the symmetry axis and
anisotropy coefficients $\epsilon$ and $\delta$ are seriously
distorted, there is some logic in the obtained azimuth $\beta =
45^\circ$. Note, this azimuth, picked by the inversion
algorithm, is equal to the azimuth of lateral velocity gradient
and corresponds to the direction of the vertical symmetry plane
which exists in both homogeneous TTI and LH isotropic models.
The parameters of inverted TTI model presented in Table~1
produce the value of the objective function ${\cal F} =
0.0013$, which corresponds to the standard deviation of
$\tilde{q}$ from $q$ [see equation~(\ref{eq02})] of only 0.26\%
relative to the measured $q|_{_{p=0}} = 0.5$~s/km at zero
horizontal slowness (Figure~\ref{fig02}b). {Thus, the
homogeneous TTI model with parameters from Table~1 accurately
fits the slowness surface obtained from VSP data in the
isotropic LH model.}

\begin{figure*}
	\centerline{
		\begin{tabular}{||l | c | c | c | c | c | c||} \hline \hline
			Model 1 & $V_0$  & $h$  & $\epsilon$ & $\delta$ & $\nu$ & $\beta$ \\
			& (km/s) & (s$^{-1})$ & & & & \\ \hline
			Correct (LH isotropic)     & 2.00 & 0.1 & 0.0  & 0.0  & --   & --   \\ \hline
			Inverted (homogeneous TTI) & 1.52 & 0.0 & 0.29 & 0.11 & 14.9 & 45.0 \\ \hline
			\hline
			\end {tabular}
		}
		\vspace*{0.5cm}
		\noindent
		{\bf Table~1.}
		Parameters of correct LH isotropic model and inverted homogeneous TTI model.
		The tilt $\nu$ and the azimuth $\beta$ (with respect to direction $x_1$
		in Figure~\ref{fig02}a) of the symmetry axis in inverted TTI model are
		given in degrees.
	\end{figure*}
	
\section{Correction for lateral heterogeneity in a single layer}

\begin{figure}[ht]
	\includegraphics[width = 2.8 in]{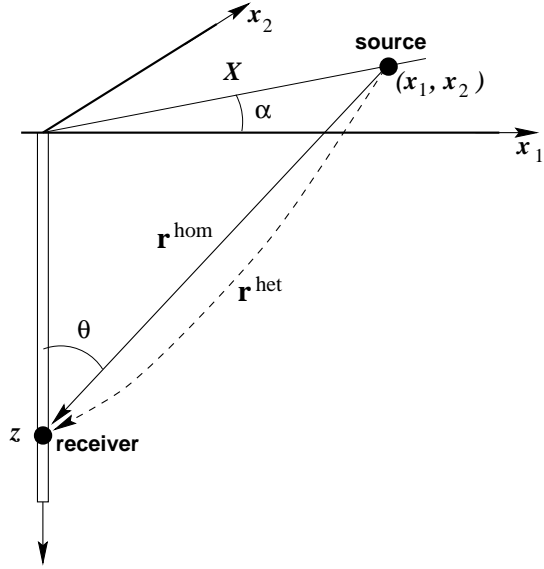}
	\caption{In deriving linearized correction for weak lateral heterogeneity,
		traveltime can be integrated along the straight ray
		${\bf r}^{\rm hom}$ (solid) in a homogeneous background model.
		The actual bending ray ${\bf r}^{\rm het}$ (dashed) does not need
		to be considered.
	}
	\label{fig03}
\end{figure}

In this section, we derive an explicit correction for lateral
heterogeneity of traveltime recorded in VSP experiment. We
consider a single, vertically homogeneous layer
(Figure~\ref{fig03}) where the group velocity $g$ can be
represented in the {\em factorized}$\,$ form:
\begin{equation}
   g(\theta, \alpha, y_1, y_2) = g^{\rm hom}(\theta, \alpha) \,
                                 f(y_1, y_2) \, ,
   \label{eq03}
\end{equation}
where $g^{\rm hom}(\theta, \alpha)$ in the group velocity in a
homogeneous anisotropic background medium. Since the ray ${\bf
r}^{\rm hom}$ between the source at $[x_1, x_2, 0]$ and the
receiver at $[0, 0, z]$ in the background medium is a straight
line, $g^{\rm hom}$ is a function of two directional angles --
the polar angle $\theta$ and the azimuth $\alpha$
(Figure~\ref{fig03}). The factor $f(y_1, y_2)$, which depends
on horizontal coordinates $y_1$ and $y_2$ along a ray,
represents lateral variation of the group velocity. We assume
that lateral heterogeneity not only factorized [for this
reason, $g$ is the product of $g^{\rm hom}$ and $f$ in
equation(\ref{eq03})] but also {\em weak}, which means that
\begin{equation}
   f(y_1, y_2) \approx 1 \, .
   \label{eq04}
\end{equation}

It is convenient to express the factor $f(y_1, y_2)$ as a
polynomial of finite power $M$
\begin{equation}
   f(y_1, y_2) \equiv 1 + \Psi(y_1, y_2) =
                      1 + \sum_{m=1}^M \sum_{\ell=0}^m \psi_{\ell,m-\ell} \,
                                       y_1^\ell \, y_2^{m-\ell} \,
   \label{eq05}
\end{equation}
with some coefficients $\psi_{\ell,m-\ell}$.
Equation~(\ref{eq04}) implies the inequality
\begin{equation}
   |\Psi(y_1, y_2)| \ll 1 \, .
   \label{eq06}
\end{equation}

We are interested in deriving the first-order correction of
traveltime due to the presence of lateral heterogeneity.
Therefore, we linearize the traveltime with respect to LH and
ignore all terms containing quadratic and higher-order
combinations of coefficients $\psi_{\ell,m-\ell}$. In the
linear approximation, the traveltime $t^{\rm LH}$ between the
source and receiver in Figure~\ref{fig03} can be calculated as
an integral along the ray ${\bf r}^{\rm hom}$ in the background
medium (e.g., Backus and Gilbert, 1969) \protect\cite{bac69}:
\begin{equation}
   t^{\rm LH}(x_1, x_2, z) =
   \frac{\sqrt{X^2 + z^2}}{X} \,
   \int_0^X \frac{d \xi}{g(\theta, \alpha, y_1(\xi), y_2(\xi))} \, ,
   \label{eq07}
\end{equation}
where
\begin{equation}
   x_1 = X \cos \alpha  \qquad {\rm and} \qquad
   x_2 = X \sin \alpha \, ;
   \label{eq08}
\end{equation}
$X$ is the offset, $\xi$ is the lateral coordinate along the
straight ray ${\bf r}^{\rm hom}$. The coordinates $y_1$ and
$y_2$ relate to $\xi$ as
\begin{equation}
   y_1 = \xi \cos \alpha  \qquad {\rm and} \qquad
   y_2 = \xi \sin \alpha \, .
   \label{eq09}
\end{equation}

Evaluating integral~(\ref{eq07}) in the linear approximation
with respect to $\psi_{\ell,m-\ell}$ yields
\begin{equation}
   t^{\rm LH}(x_1, x_2, z) = t^{\rm hom}(x_1, x_2, z) \,
                              {\cal H}(x_1, x_2) \, ,
   \label{eq10}
\end{equation}
where
\begin{equation}
   t^{\rm hom}(x_1, x_2, z) =
   \frac{\sqrt{X^2 + z^2}}{g^{\rm hom}(\theta, \alpha)} \,
   \label{eq11}
\end{equation}
is the traveltime in the background medium, and the factor
\begin{eqnarray}
   \label{eq12}
   {\cal H}(x_1, x_2) \equiv 1 + H(x_1, x_2) = 1 - \sum_{m=1}^M \frac{1}{m+1} \, \sum_{\ell=0}^m \psi_{\ell,m-\ell} \, x_1^\ell \, x_2^{m-\ell} \,
\end{eqnarray}
will be called the heterogeneity factor. Comparing
equations~(\ref{eq05}) and~(\ref{eq12}), we conclude that,
according to inequality~(\ref{eq06}), the magnitude of
polynomial $H(x_1, x_2)$ has to be small, i.e.,
\begin{equation}
   |H(x_1, x_2)| \ll 1 \, .
   \label{eq13}
\end{equation}

\subsection{Accounting for lateral heterogeneity}

Equation~(\ref{eq10}), the main result of previous section,
holds in arbitrarily anisotropic, factorized, weakly LH media.
Here, we use it to infer the velocity function specified by
coefficients $\psi_{\ell,m-\ell}$ from measured traveltime
$t^{\rm LH}$ and its derivatives
\begin{eqnarray}
   \label{eq14}
   p_i^{\rm LH} \equiv \frac{\partial t^{\rm LH}}{\partial x_i}
   \qquad \text{and} \qquad q^{\rm LH} \equiv \frac{\partial t^{\rm LH}}{\partial z} \, ,
   \qquad (i=1,2).
\end{eqnarray}

Substituting equation~(\ref{eq10}) into equations~(\ref{eq14})
yields
\begin{eqnarray}
   \label{eq15}
   p_i^{\rm LH} = p_i^{\rm hom} \, {\cal H} +
                   t^{\rm hom} \, \frac{\partial {\cal H}}{\partial x_i}
   \qquad \text{and} \qquad q^{\rm LH} = q^{\rm hom} \, {\cal H} \, ,
\end{eqnarray}
where, by definition,
\begin{equation}
   p_i^{\rm hom} \equiv \frac{\partial t^{\rm hom}}{\partial x_i}
   \qquad \text{and} \qquad
   q^{\rm hom} \equiv \frac{\partial t^{\rm hom}}{\partial z} \, .
   \label{eq16}
\end{equation}
Traveltime $t^{\rm hom}$ and components of the slowness vector
in homogeneous anisotropic media relate as
\begin{equation}
   p_1^{\rm hom} \, x_1 + p_2^{\rm hom} \, x_2 + q^{\rm hom} \, z
   = t^{\rm hom} \, .
   \label{eq17}
\end{equation}
This equation follows from the relation between the slowness
and the group velocity vectors (e.g., Musgrave, 1970)
\protect\cite{mus}
\begin{equation}
   p_1^{\rm hom} \, g_1^{\rm hom} + p_2^{\rm hom} \, g_2^{\rm hom} +
   q^{\rm hom} \, g_3^{\rm hom} = 1 \, .
   \label{eq18}
\end{equation}

Using the measured quantities $p_i^{\rm LH}$, $q^{\rm LH}$, and
$t^{\rm LH}$, we construct the {\em known} quantity
\begin{equation}
   {\cal R} = \frac{1}{t^{\rm LH}} \,
   \left( p_1^{\rm LH} \, x_1 + p_2^{\rm LH} \, x_2 + q^{\rm LH} \, z
   \right) - 1 \, .
   \label{eq19}
\end{equation}
Taking into account equations~(\ref{eq10}), (\ref{eq12}),
(\ref{eq15}) and~(\ref{eq17}), equation~(\ref{eq19}) can be
reduced to
\begin{equation}
   {\cal R} \, (1 + H) =
   \frac{\partial H}{\partial x_1} \, x_1 +
   \frac{\partial H}{\partial x_2} \, x_2 \, .
   \label{eq20}
\end{equation}
This equation allows us to find $H$ or the coefficients
$\psi_{\ell,m-\ell}$ [see equation~(\ref{eq12})] defining
velocity heterogeneity. We express ${\cal R}$ as a polynomial
\begin{equation}
   {\cal R}(x_1, x_2) = \sum_{m=1}^M \sum_{\ell=0}^m
                        R_{\ell,m-\ell} \, x_1^\ell \, x_2^{m-\ell} \,
   \label{eq21}
\end{equation}
and note that ${\cal R}(x_1, x_2)$ does not have a constant
(i.e., independent on $x_i$) term because there is no such a
term in the right-hand side of equation~(\ref{eq20}). Also, due
to inequality~(\ref{eq13}), the magnitude of polynomial ${\cal
R}$ is small. Substituting equations~(\ref{eq12})
and~(\ref{eq21}) into equation~(\ref{eq20}) and solving this
equation in the linear approximation for the coefficients
$\psi_{\ell,m-\ell}$ yields the set of explicit relations
\begin{equation}
   \psi_{\ell,m-\ell} = - \frac{m+1}{m} \, R_{\ell,m-\ell} \, .
   \label{eq22}
\end{equation}

Thus, we have obtained the following algorithm to remove the
influence of LH on VSP data: \vspace{-0.2cm}
\begin{itemize}
\item construct the function ${\cal R}$ given by
    equation~(\ref{eq19}) from the measured quantities
    $p_i^{\rm LH}$, $q^{\rm LH}$, and $t^{\rm LH}$;
\item represent ${\cal R}(x_1, x_2)$ as the
    polynomial~(\ref{eq21}) and, using
    equation~(\ref{eq22}), compute the coefficients $\psi$
    of the function $f$ [equation~(\ref{eq05})] describing
    lateral velocity heterogeneity;
\item build the heterogeneity factor ${\cal H}$
    [equation~(\ref{eq12})] and reconstruct the slowness
    components $p_i^{\rm hom}$ and $q^{\rm hom}$ in the
    homogeneous background medium using
    equations~(\ref{eq10}) and (\ref{eq15}).
\end{itemize}

\begin{figure}[ht]
	\includegraphics[width = 5.5 in]{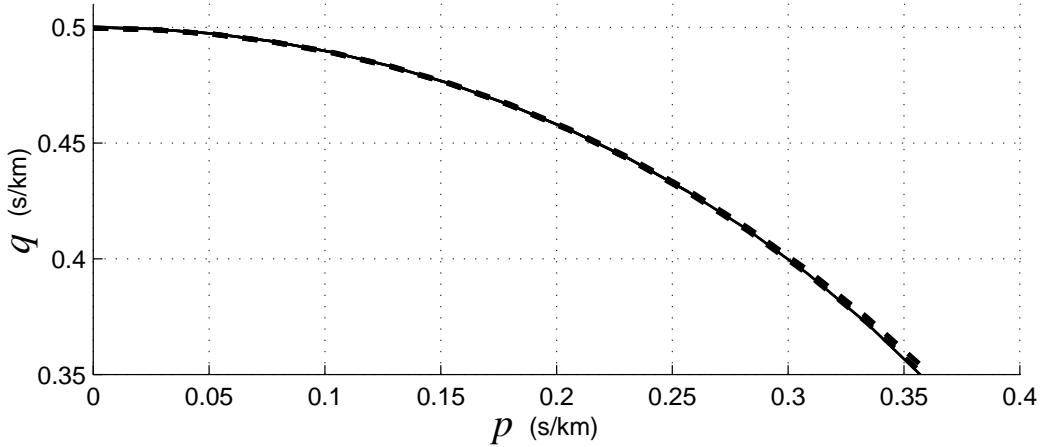}
	\caption{Cross-sections of the slowness surface
		$q(p \cos \alpha,$ $p \sin \alpha)$ at azimuths
		$\alpha = 0^\circ, \, 45^\circ, \, 90^\circ, \, 135^\circ$, and
		$180^\circ$ reconstructed after accounting for lateral
		heterogeneity (overlapping dashed lines) and cross-section of the
		correct isotropic slowness surface (solid line). Parameters of the
		correct model are given in Table~1.
	}
	\label{fig04}
\end{figure}

\section{Numerical examples}

Here, we present several
numerical examples to test the technique developed in the
previous section. We begin with the isotropic LH model from Table~1.
Our algorithm accurately
reconstructed lateral velocity gradient and produced the
slowness surface which cross-sections are shown in
Figure~\ref{fig04} (compare it with Figure~\ref{fig02}b). The
cross-sections at different azimuths overlap each other and the
correct isotropic cross-section. There is small deviation at
large values of $p$ where LH influences traveltimes stronger.
We inverted the reconstructed slowness surface and obtained the
following set of parameters of TTI medium: $V_0 = 1.99$~km/s,
$\epsilon = 0$, $\delta = 0.01$, $\nu = 41.7^\circ$, and $\beta
= 45.0^\circ$ (compare these parameters with the correct
parameters in Table~1). Interestingly, that the inversion
algorithm again picked up the azimuth $\beta = 45^\circ$ of the
vertical symmetry plane of the model.


An alternative approach to account for lateral velocity variations is to replace the horizontal slowness components with the $P$-wave polarization directions \protect\cite{mat}. The authors of paper \protect\cite{mat}, published later, derived the $P$-wave
slowness on polarization dependence, established its governing parameters, and estimated those parameters from a $P$-wave VSP
data set acquired in the deep-water Gulf of Mexico (see Figure 1 in
Grechka and Mateeva \protect\cite{mat} and references therein).

\begin{figure*}
	\centerline{
		\begin{tabular}{||l | c | c | c | c | c||} \hline \hline
			Model 2  & $V_0$ (km/s) & $\epsilon$ & $\delta$ & $\nu$ & $\beta$ \\ \hline
			Correct  & $2.00 \, (1 + 0.100 x_1 + 0.030 x_2 + 0.010 x_1^2)$ &
			0.250 & 0.150 & 30.0 & --90.0   \\ \hline
			Inverted & $1.98 \, (1 + 0.100 x_1 + 0.029 x_2 + 0.003 x_1^2)$ &
			0.252 & 0.149 & 29.6 & --90.0 \\ \hline
			\hline
			\end {tabular}
		}
		\vspace*{0.5cm}
		\noindent
		{\bf Table~2.}
		Parameters of correct and inverted LH TTI models.
		The tilt $\nu$ and the azimuth $\beta$ of the symmetry axis are
		given in degrees.
\end{figure*}
	
\begin{figure}[ht]
\includegraphics[width = 4.3 in]{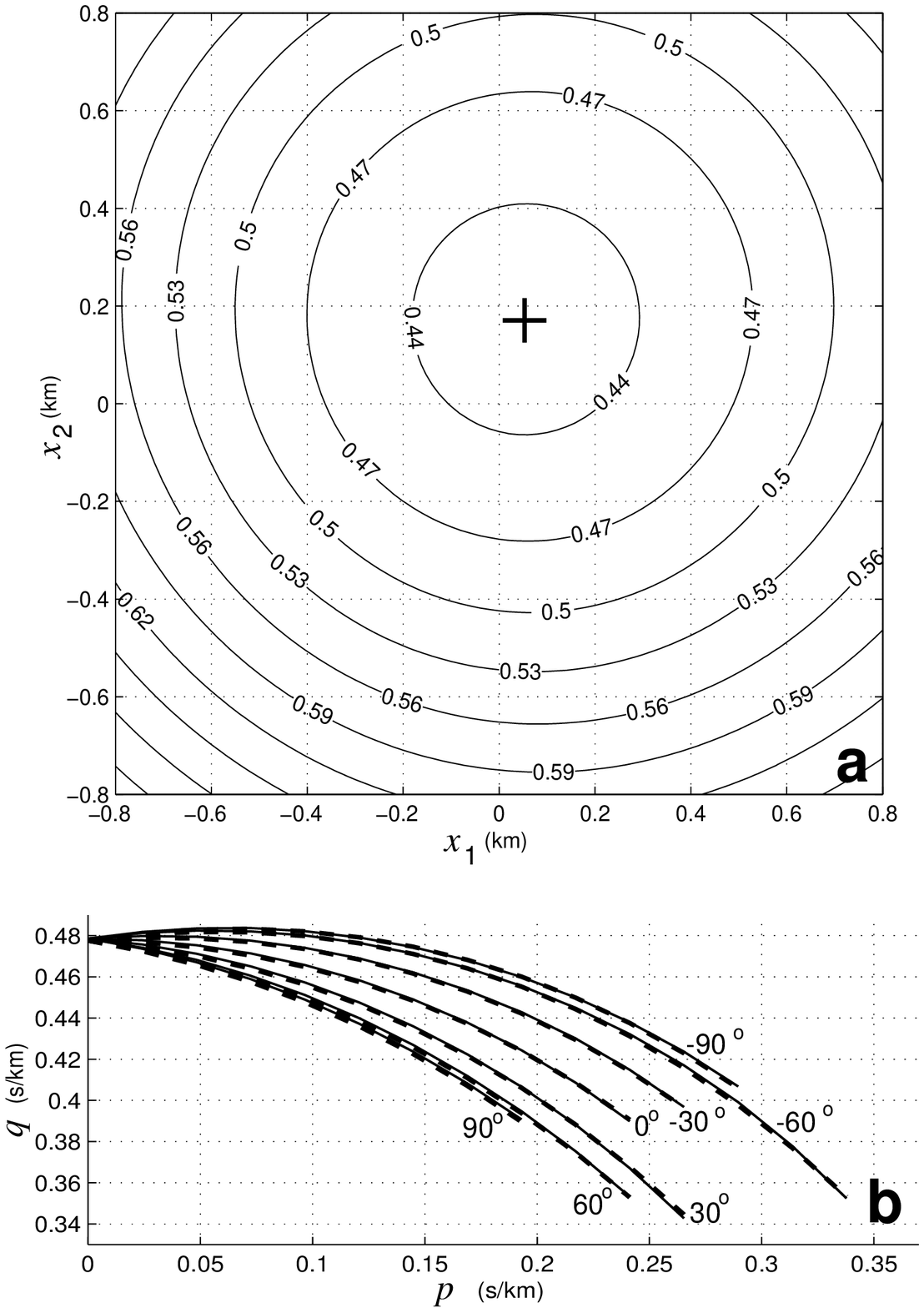}
\caption{(a)~Contours (in s) of traveltime $t^{\rm LH}(x_1, x_2)$
         recorded at depth $z = 0.94$~km.
         The ``{\bf +}'' sign indicates the position of traveltime
         minimum.
         (b)~Cross-sections of the slowness surface
         $q^{\rm hom}(p_1^{\rm hom}, p_2^{\rm hom})$, where
         $p_1^{\rm hom} = p \cos \alpha$ and $p_2^{\rm hom} = p \sin \alpha$,
         at azimuths
         $\alpha = -90^\circ, \, -60^\circ, \, -30^\circ,$ $0^\circ, \,
         30^\circ, \, 60^\circ$, and $90^\circ$ reconstructed after
         accounting for lateral heterogeneity (dashed) and
         cross-sections of the correct TTI slowness surface (solid).
        }
\label{fig05}
\end{figure}

In the second test, we estimate parameters of LH TTI model
in Table~2. This time, the shift of traveltime minimum (marked
with the plus in Figure~\ref{fig05}a) from the coordinate
origin is due to the influence of both anisotropy without
horizontal symmetry plane and lateral heterogeneity. The
technique described in the previous section has been able to
separate these two influences and resulted in the
cross-sections of the slowness surface shown in
Figure~\ref{fig05}b with dashed lines. Their small deviations
from the correct cross-sections (solid lines in
Figure~\ref{fig05}b) can be explained by the approximations
made with respect to lateral heterogeneity. {The
	inverted anisotropic parameters are given in Table~2. Although
	there are some errors in the estimated parameters, the
	algorithm has performed fairy well.}

\begin{figure}
	\centerline{
		\begin{tabular}{||l | c | c | c | c||} \hline \hline
			Model 3  & $\epsilon$ & $\delta$ & $\nu$ & $\beta$ \\ \hline
			Correct  & 0.250 & variable & variable & --90.0   \\ \hline
			Inverted & 0.251 & 0.158    & 30.1     & --90.2   \\ \hline
			\hline
		\end {tabular}
		}
		\vspace*{0.5cm}
		\noindent
		{\bf Table~3.}
		Parameters of correct and inverted LH TTI models. The lateral variation
		of velocity $V_0$ is the same as that in Model 2 (Table 2). The anisotropy
		coefficient $\delta$ changes in the direction $x_2$ as
		$\delta = 0.15 (1.0 + 0.2 x_2)$.
		The lateral variation of the tilt $\nu$ is
		$\nu = \arcsin [\, 0.5 + 0.125 (x_1-x_2)\,]$.
\end{figure}
	
In our last test in this section, we examine what happens if
the assumption of factorized anisotropy is violated, and both
the velocity $V_0$ and anisotropic coefficients change
laterally. The model parameters are shown in Table~3. Three
quantities -- $V_0$, $\delta$, and $\nu$ -- vary. While
variation of $\delta$ is relatively small (it changes from
0.126 to 0.174 with the mean value equal to 0.15), the
variation of the tilt $\nu$ is more pronounced. The tilt varies
from $\nu = 17.1^\circ$ to $\nu = 40.1^\circ$, the mean tilt is
$28.6^\circ$. Again, we apply the same technique to estimate
the influence of LH, remove it, and infer parameters of TTI
medium assuming that {\em anisotropy is factorized}. The
inversion results are presented in Table~3. We did not
reconstruct lateral variations of anisotropic coefficients,
instead, we obtained their single values which happened to be
close to the corresponding correct mean values. {This
suggests that the assumption of factorized anisotropy is not
that important, and obtaining some reasonable estimates of
anisotropic parameters, close to their mean values, can be
expected.}

\section{Ambiguity between vertical heterogeneity VI and lateral heterogeneity LH}

{Since vertical inhomogeneity (VI() is usually stronger
than lateral heterogeneity LH}, it is important the develop a
technique that would allow one to account for LH {\em in the
presence of VI}. {First, we show that the required data
should contain traveltimes recorded at a sufficiently dense set
of downhole receivers, otherwise it is impossible to separate
the influences of LH and VI on VSP data even though there are
measurements of the vertical velocity along a borehole}.

\begin{figure}[ht]
\includegraphics[width = 3.0 in]{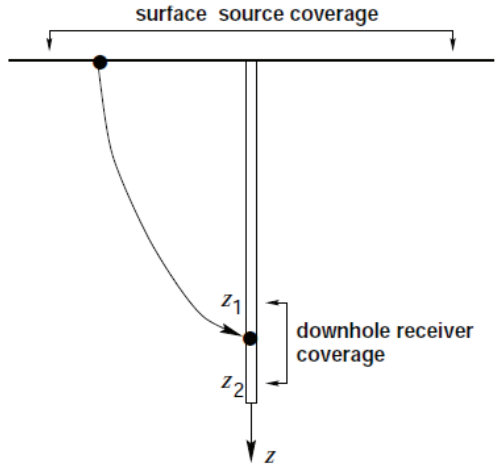}
\caption{VSP data with only
         partial coverage of downhole receivers do not allow one to
         distinguish between lateral and vertical heterogeneity.
        }
\label{fig06}
\end{figure}

Suppose we record traveltimes in the VSP geometry
(Figure~\ref{fig06}), where surface sources have a large
aperture whereas downhole receivers cover only a relatively
small range of depths $z_1 \le z \le z_2$. We assume that the
depth coverage is just sufficient to calculate the derivative
$\partial t / \partial z$ [i.e., $(z_2 - z_1)/z_2 \ll 1$] and
find the vertical slowness $q$ at depth $z$. As a test, we
examine what can be reconstructed from the traveltime $t^{\rm
VI}$ and the vertical slowness $q^{\rm VI}$ in elliptically
anisotropic VTI model with horizontal velocity changing
linearly with depth (Table~4).
%
Figure~\ref{fig07}a shows traveltime $t^{\rm VI}$ (dotted line)
calculated using an exact equation in elliptically anisotropic
VI medium with parameters from Table~4. The traveltime $t^{\rm
VI}$ corresponds to the receiver depth $z = 1$~km, where
Thomsen's anisotropic coefficients $\epsilon$ and $\delta$
reach 0.28 indicating significant anisotropy. Despite that,
traveltime $t^{\rm VI}$ virtually coincides with traveltime
$t^{\rm LH}$ (solid line) computed numerically in LH medium
with non-elliptical anisotropy (parameters of this medium are
given in Table~4; the function $V_0(x)$ is shown in
Figure~\ref{fig07}b). The maximum difference between
traveltimes $t^{\rm VI}$ and $t^{\rm LH}$ is only 0.3\%,
illustrating a high accuracy of the approximations made and also
showing that both traveltimes are indistinguishable.

\begin{figure*}
\centerline{
\begin{tabular}{||l | c | c | c||} \hline \hline
Model 4       & Vertical velocity & $\epsilon$ & $\delta$ \\ \hline
              &                   &            & \\
Correct (VI)  & $V_0 = {\rm const}$
              & $kz + \dfrac{(kz)^2}{2}$
              & $kz + \dfrac{(kz)^2}{2}$ \\
              &                   &            & \\ \hline
              &                   &            & \\
Inverted (LH) & $V_0 \left( 1 + \dfrac{3k}{4z} \, x^2 -
                                \dfrac{5k}{8z^3} \, x^4 \right) $
              & $\dfrac{kz}{8}$
              & $\dfrac{kz}{4}$ \\
              &                   &            & \\ \hline
\hline
\end {tabular}
}
\vspace*{0.5cm}
\noindent
{\bf Table~4.}
Parameters of correct and inverted VTI models.
The correct vertically inhomogeneous model is elliptically anisotropic.
It has constant vertical velocity $V_0$ and depth-varying anisotropic
coefficients $\epsilon = \delta$.
The inverted model is laterally heterogeneous.
Its vertical velocity changes laterally and anisotropy is not elliptical
($\epsilon \ne \delta$).
\end{figure*}

\begin{figure}[ht]
\includegraphics[width = 4.0 in]{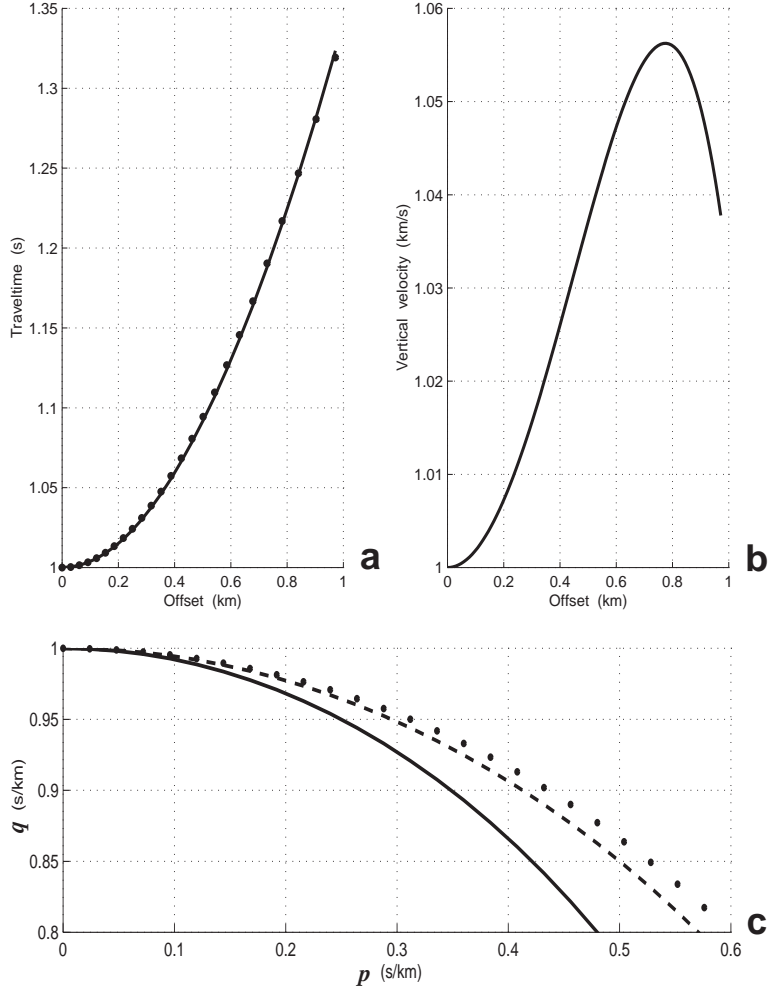}
\caption{(a) Traveltimes in VTI models from Table~4:
         $t^{\rm VI}$ (dotted) calculated under the assumption of VI
         using exact expression, and
         $t^{\rm LH}$ (solid) computed numerically in the LH model.
         (b) Lateral variation of the vertical velocity $V_0(x)$ given in
         Table~4.
         (c) Slowness curves:
         in VI model (solid) and
         in LH model (dashed).
         The dotted line indicates a circle.
         Model parameters: $V_0 = 1$~km/s, $k = 0.25$~km$^{-1}$.
         The depth of downhole receiver $z = 1$~km.
        }
\label{fig07}
\end{figure}

The results of reconstruction of the slownesses are presented in
Figure~\ref{fig07}c. While the solid line represents elliptical
dependence that has been estimated assuming vertically
inhomogeneous model, the dashed line (close to the circular
dotted line) shows quite different function obtained under the
assumption that the model is laterally heterogeneous. Since
both models fit the traveltime and the vertical slowness
measured at single depth $z$, it is impossible to infer the
type of heterogeneity -- vertical versus lateral -- from the
data.

\subsection{Correcting for LH in the presence of VI}

{To distinguish between lateral and vertical
heterogeneity, it is necessary to acquire VSP data with such a
dense sampling along a borehole that the medium between
adjacent receivers can be considered vertically homogeneous.}
Then, the problem reduces to the sequence of parameter
estimation steps for each interval between adjacent receivers
(or groups of receivers; the groups are needed to estimate the
vertical slowness component) followed by downward propagation
of traveltimes within the intervals.

\begin{figure}[htp]
\includegraphics[width = 2.5 in]{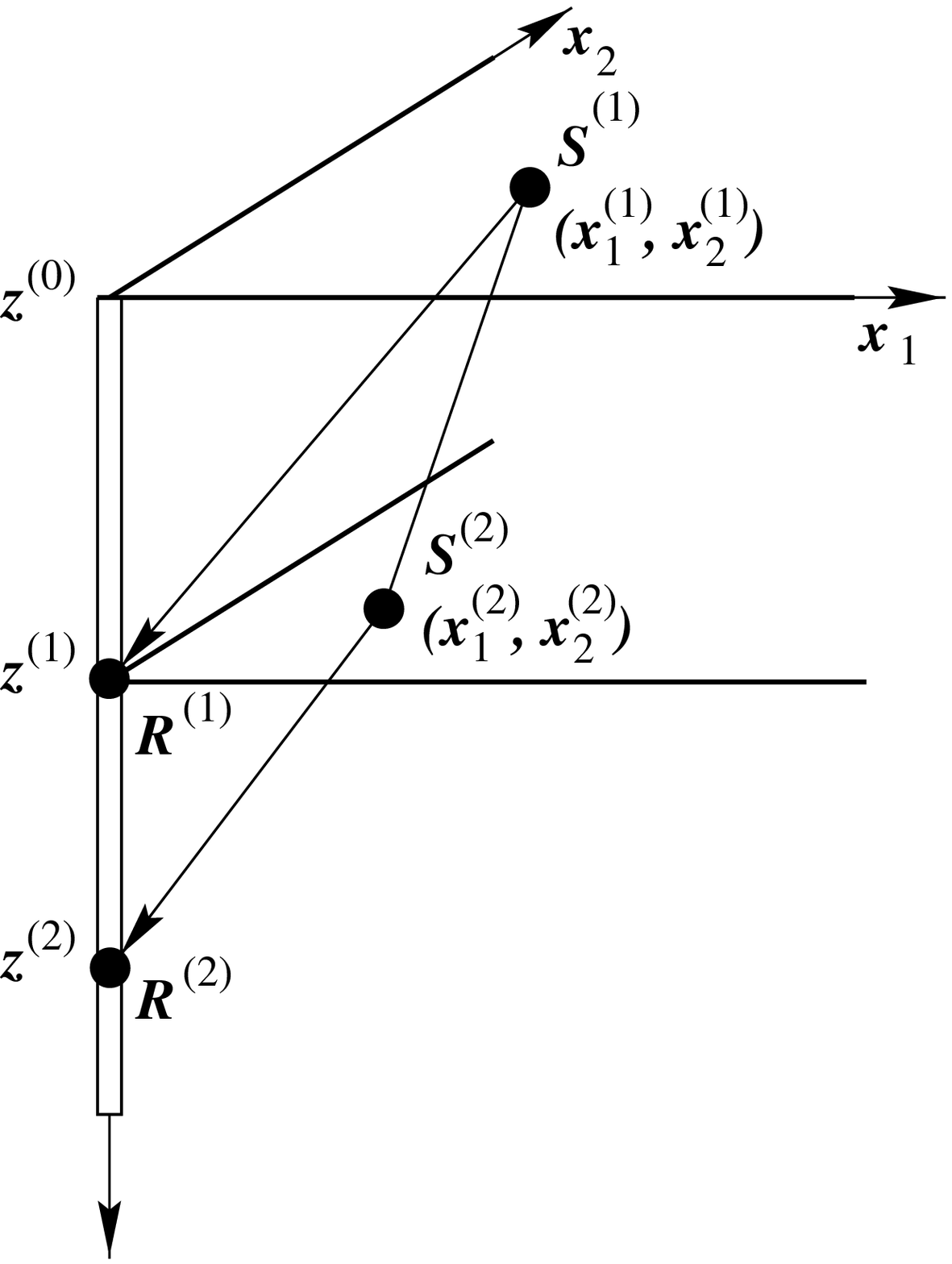}
\caption{Accounting for LH in the presence of VI can be done in two steps.
         First, we remove LH and estimate anisotropic parameters in the
         interval $[z^{(0)}, z^{(1)}]$ above receiver $R^{(1)}$ applying the
         technique developed for a single layer.
         Second, we use obtained parameters to map the traveltime
         measured between the source $S^{(1)}$ and receiver $R^{(2)}$ onto
         the depth level $z^{(1)}$. Then, we repeat the first step
         to find parameters in the interval $[z^{(1)}, z^{(2)}]$ between
         receivers $R^{(1)}$ and $R^{(2)}$.
        }
\label{fig08}
\end{figure}

The algorithm is illustrated in Figure~\ref{fig08}. Let us
suppose that the data contain two sets of traveltimes $t^{\rm
LH}(n)$ and vertical slowness components $q^{\rm LH,(n)}$
$(n=1,2)$ recorded between the sources $S^{(1)}$ at the surface
$z^{(0)}$ and receivers $R^{(n)}$ at depths $z^{(n)}$. The
index $n$ denotes traveltimes measured by receiver $R^{(n)}$;
$n$ in the superscript refers to the interval quantities.  We
assume that the distances $z^{(n)} - z^{(n-1)}$ are
sufficiently small so that the medium at each interval can be
considered vertically homogeneous. We can apply the above
described method to the measured traveltimes $t^{\rm LH}(1) =
t^{\rm LH,(1)}$ and vertical slowness components $q^{\rm
LH,(1)}$ to estimate velocity heterogeneity $f^{(1)}$
[equation~(\ref{eq03})], the slowness vectors ${\bf p}^{\rm
hom,(1)} = [p_1^{\rm hom,(1)}, p_2^{\rm hom,(1)}, q^{\rm
hom,(1)}]$, and vector of anisotropic parameters $\chi^{(1)} =
[V_0^{(1)}, \epsilon^{(1)}, \delta^{(1)}, \nu^{(1)},
\beta^{(1)}]$ in the first interval between the levels
$z^{(0)}$ and $z^{(1)}$.

Next, we take the measured traveltimes $t^{\rm LH}(2)$ between
the sources $S^{(1)}$ located at $[x_1^{(1)}, x_2^{(1)},
z^{(0)}]$ and the receiver $R^{(2)}$ at $[0, 0, z^{(2)}]$
(Figure~\ref{fig08}) and find the horizontal slowness
components $p_i^{\rm LH,(1)} = \partial t^{\rm LH}(2) /
\partial x_i$ at level $z^{(0)}$. Using the assumption of
factorized anisotropy, the horizontal slowness components in
the background medium can be computed as
\begin{equation}
   p_i^{\rm hom,(1)} = p_i^{\rm LH,(1)} \, f^{(1)} \, , \qquad \qquad
                       (i=1,2) \, .
   \label{eq23}
\end{equation}
Then, the value of $q^{\rm hom,(1)}$ is obtained from the
already known vectors ${\bf p}^{\rm hom,(1)}$. This allows us
to shoot rays from the sources $S^{(1)}$ downward and find
lateral coordinates of the points $S^{(2)}$ at the level
$z^{(1)}$:
\begin{eqnarray}
   x_i^{(2)} = x_i^{(1)} - \frac{g_i^{\rm hom,(1)}}{g_3^{\rm hom,(1)}} \, \left(z^{(1)} - z^{(0)} \right)
   = x_i^{(1)} + q_{,i}^{\rm hom,(1)} \, \left(z^{(1)} - z^{(0)} \right) \, ,
   \label{eq24}
\end{eqnarray}
where $q_{,i}^{\rm hom,(1)} \equiv
\partial q^{\rm hom,(1)} / \partial p_i^{\rm hom,(1)}$, and
the relation $q_{,i}^{\rm hom,(1)} = - g_i^{\rm hom,(1)} /
g_3^{\rm hom,(1)}$ is obtained by differentiating
equation~(\ref{eq18}).

The traveltimes $t^{\rm LH,(1)}$ in the depth interval
$[z^{(0)}, z^{(1)}]$ between the points $S^{(1)}$ at
$[x_1^{(1)}, x_2^{(1)}, z^{(0)}]$ and $S^{(2)}$ at $[x_1^{(2)},
x_2^{(2)}, z^{(1)}]$ can be calculated based on
equation~(\ref{eq07}) under the assumption of weak lateral
heterogeneity:
\begin{eqnarray}
   t^{\rm LH,(1)}  = \sqrt{1 + \frac{\Delta z^2}{\Delta X^2} } \int_0^{\Delta X} \frac{d \xi} {g^{\rm hom,(1)} \, f^{(1)}(y_1(\xi), y_2(\xi))} \, ,
   \label{eq25}
\end{eqnarray}
where
\[ \Delta X = \left[ \left(x_1^{(2)} - x_1^{(1)} \right)^2 +
                     \left(x_2^{(2)} - x_2^{(1)} \right)^2
              \right]^{\frac{1}{2}},
\]
\[              \Delta z = z^{(1)} - z^{(0)}, \]
\[  y_1 = x_1^{(1)} + \xi \cos \alpha, \]
\[  y_2 = x_2^{(1)} + \xi \sin \alpha, \]
and
\[
   \tan \alpha = \frac{x_2^{(2)} - x_2^{(1)}}
                      {x_1^{(2)} - x_1^{(1)}} \, .
\]
The group velocity $g^{\rm hom,(1)}$ in equation~(\ref{eq25})
is computed using the already found components of the slowness
vector ${\bf p}^{\rm hom,(1)}$ and their derivatives obtained from equation~(\ref{eq18}),
\begin{equation}
   \label{eq26}
   g^{\rm hom,(1)} = \sqrt{ \sum_{i=1}^3 \left( g_i^{\rm hom,(1)} \right)^2 }
                   = \frac{\sqrt{1 + \left( q_{,1}^{\rm hom,(1)} \right)^2 +
                         \left( q_{,2}^{\rm hom,(1)} \right)^2 } }
              {q^{\rm hom,(1)} - q_{,1}^{\rm hom,(1)} p_1^{\rm hom,(1)} -
                                 q_{,2}^{\rm hom,(1)} p_2^{\rm hom,(1)} }.
\end{equation}

Equations~(\ref{eq24}) and~(\ref{eq25})
allow us to continue the traveltimes measured at level
$z^{(0)}$ onto level $z^{(1)}$. The traveltime $t^{\rm LH}(2)$
recorded at $[x_1^{(1)}, x_2^{(1)}, z^{(0)}]$ is mapped onto
traveltime
\begin{equation}
   t^{\rm LH,(2)} = t^{\rm LH}(2) - t^{\rm LH,(1)}
   \label{eq27}
\end{equation}
at $[x_1^{(2)}, x_2^{(2)}, z^{(1)}]$. Then, using the measured
vertical slowness $q^{\rm LH,(2)}$, we perform the parameter
estimation step for the second interval between the depths
$z^{(1)}$ and $z^{(2)}$, and find the vector $\chi^{(2)} =
[V_0^{(2)}, \epsilon^{(2)}, \delta^{(2)}, \nu^{(2)},
\beta^{(2)}]$.

{Although the entire process can be repeated as many
times as desirable, it may lead to increasing errors in
estimated parameters as we go from one interval to another.}
The main source of errors are inaccuracies in the derivatives
$q_{,i}^{{\rm hom,}(n)} =
\partial q^{{\rm hom,}(n)} / \partial p_i^{{\rm hom,}(n)}$ ($n$ is the number
of an interval), determining the directions of downward
propagating rays in equation~(\ref{eq24}). The slowness
components are the derivatives of measured traveltimes and,
therefore, contain a higher level of errors than the
traveltimes themselves. Taking the derivative $\partial q^{{\rm
hom,}(n)} / \partial p_i^{{\rm hom,}(n)}$, we differentiate the
traveltimes twice, further amplifying those errors.
The correction for LH is also a potential source of errors
because it is the approximation designed for weak lateral
heterogeneity. Correcting the data for LH in each interval, we
introduce some errors. These errors are cascaded and may grow
as we go deeper. Another inherent limitation of the technique
is evident from equation~(\ref{eq27}): when the interval
traveltime $t^{\rm LH,(n)}$ is small, its relative error is
large. Therefore, parameters estimated for thinner intervals may
be associated with greater errors. Overall, if the described
procedure is applied to a large number of relatively thin
intervals, the reliability of inverted anisotropic parameters
may be questionable.

\begin{figure}
\centerline{
\begin{tabular}{||c | c | c | c | c | c||} \hline \hline
Layer & $V_0$ & $\epsilon$ & $\delta$ & $\nu$ & $\beta$ \\ \hline
\multicolumn{6}{  c  }{\textbf{Correct model parameters}} \\ \hline
1     & 2.00  & 0.15       & 0.10     & 70.0  &  0.0 \\
2     & 2.30  & 0.20       & 0.15     & 60.0  & 20.0 \\
3     & 2.50  & 0.25       & 0.15     & 50.0  & 40.0 \\ \hline \hline
\multicolumn{6}{  c  }{\textbf{Inverted parameters}} \\
\multicolumn{6}{c}{LH is accounted for} \\ \hline
1     & 1.98  & 0.16       & 0.10     & 70.4  &  0.0 \\
2     & 2.25  & 0.24       & 0.16     & 58.0  & 19.6 \\
3     & 2.47  & 0.28       & 0.18     & 49.1  & 37.9 \\ \hline
\multicolumn{6}{c}{LH is ignored} \\ \hline
1     & 2.02  & 0.13       & 0.11     & 90.0  &  0.0 \\
2     & 2.34  & 0.14       & 0.07     & 65.8  & 29.9 \\
3     & 2.55  & 0.19       & 0.04     & 56.7  & 58.1 \\ \hline
\hline
\end {tabular}
}
\vspace*{0.5cm}
\noindent
{\bf Table~5.}
Parameters of correct and inverted three-layered models.
The layer thicknesses are 0.95, 0.40, and 0.10~km.
Lateral velocity variations in the layers are linear:
$f^{(1)} = 1 + 0.05 x_1$, $f^{(2)} = 1 - 0.04 x_1 - 0.02 x_2$, and
$f^{(3)} = 1 + 0.03 x_2$. Velocity $V_0$ is given in km/s, the tilt
$\nu$ and azimuth $\beta$ of the symmetry axis -- in degrees.
\end{figure}

\begin{figure}[htp]
\includegraphics[width = 5.5 in]{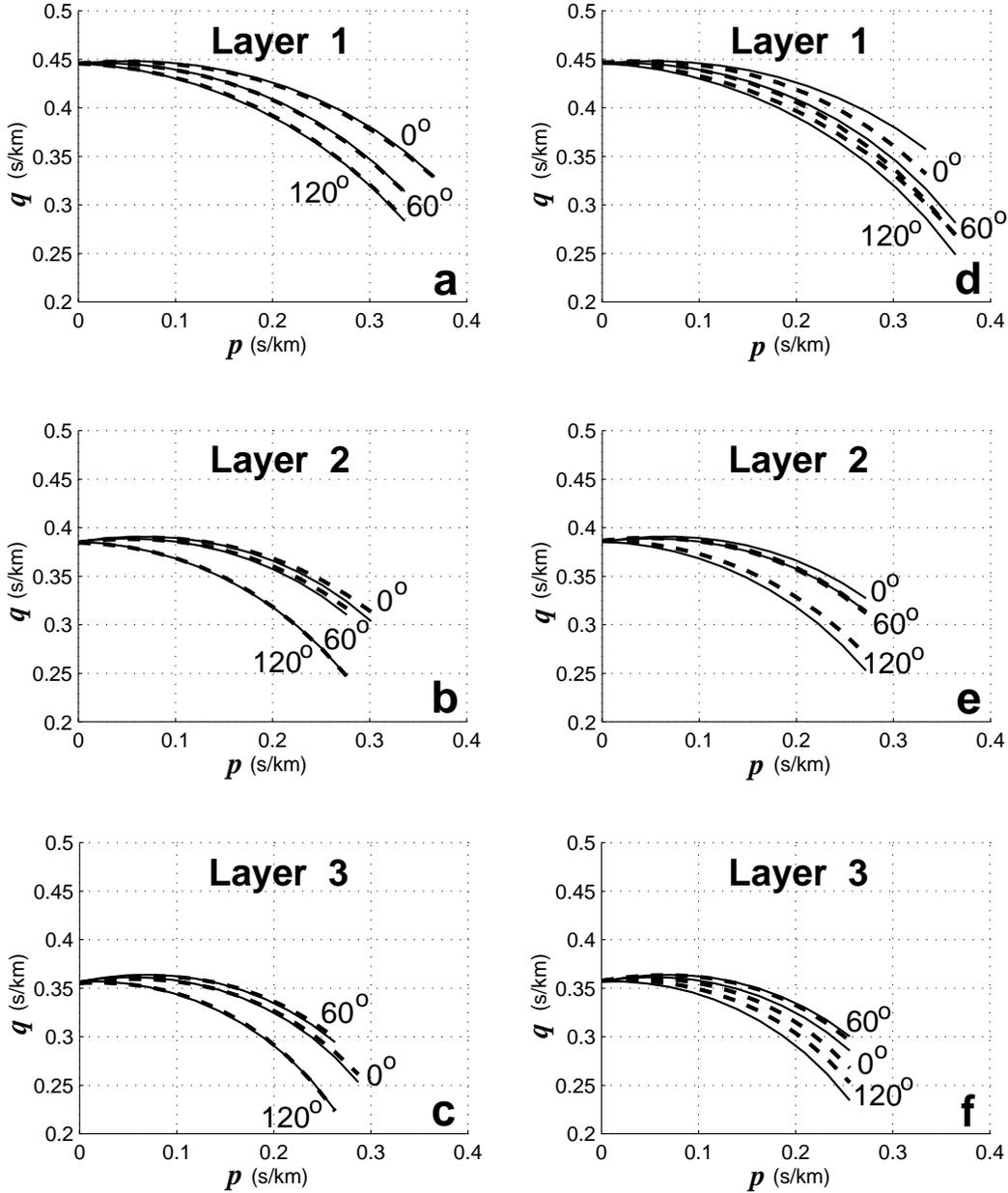}
\caption{Cross-sections of the correct slowness surfaces at
azimuths
         $\alpha = 0^\circ$, $60^\circ$, and $120^\circ$ (solid) in
         the three-layer LH VTI model given in Table~5 and the reconstructed
         cross-sections (dashed).
         Lateral heterogeneity has been estimated and removed
         in (a), (b), (c), and ignored in (d), (e), (f).
         }
\label{fig09}
\end{figure}

To illustrate the performance of correction for LH in layered
media, we invert parameters of a three-layer LH TTI model. The
correct model parameters are given in Table~5.
Figure~\ref{fig09}a--\ref{fig09}c shows the cross-sections of
the slowness surfaces in the model layers after correcting for
LH (dashed). There are some deviations of the reconstructed
cross-sections from the correct ones (solid). These deviations
lead to errors in inverted parameters of TTI layers (Table~5).
The errors, however, are relatively small and do not exceed
0.04 for anisotropic coefficients $\epsilon$ and $\delta$ and
$2.1^\circ$ for the tilt $\nu$ of the symmetry axis and its
azimuth $\beta$. For comparison, we show the cross-sections
obtained ignoring the influence of LH
(Figure~\ref{fig09}d -- \ref{fig09}f). Clearly, the difference
between the correct and reconstructed slowness surfaces became
more significant. As a result, the inverted anisotropy
parameters (Table~5) contain greater errors. The errors in the
values of $\epsilon$ and $\delta$ reach 0.11, and in $\nu$ and
$\beta$ -- $20^\circ$. This example illustrates that, although
the developed correction for LH does not produce perfect
inversion results, the errors in estimated interval parameters
are smaller than those obtained when LH is ignored.


\section{Discussion}
Lateral heterogeneity may introduce substantial errors in
anisotropic parameters inverted from multi-azimuth walkaway VSP
data. We have shown that the presence of LH can be identified
from the VSP data and developed the procedure to remove its
influence on estimated anisotropic parameters. Our technique is
based on the relation $t = ({\bf p} \cdot {\bf x})$ between
traveltime $t$ in a homogeneous arbitrary anisotropic medium,
the slowness vector {\bf p}, and the difference {\bf x} between
coordinates of a source and a receiver. VSP geometry allows one
to measure all quantities in this equation directly. Thus, we
can explicitly check if the data satisfy the hypothesis of
homogeneity. If they do not, there is a choice of attributing
the deviation of $t$ from $({\bf p} \cdot {\bf x})$ to either
vertical or lateral heterogeneity. We have shown that the
same data may correspond to heterogeneity of either type.
Therefore, it appeared that the only way to separate the two
kinds of heterogeneity in VSP geometry is to estimate vertical
inhomogeneity explicitly, using the data obtained at a
sufficiently dense set of downhole receivers. We have developed
a procedure to account for LH in vertically varying media.

The important assumption we have made is that anisotropy is
factorized within each depth interval. The physical meaning of
this assumption is that velocity heterogeneity makes greater
contribution to recorded traveltimes than variations in
anisotropic coefficients. Based on that, we ignored these
variations. The assumption of factorized anisotropy allowed us
to separate the influences of lateral heterogeneity and
anisotropy on the data. In fact, we effectively reduced the
number of parameters which determine the traveltimes in VSP
geometry and, thus, removed the otherwise existing null-space
of our inverse problem. Note that, after LH has been accounted
for, all five parameters specifying $P$-wave kinematics in
transversely isotropic media with a tilted axis of symmetry --
$V_0, \, \epsilon, \, \delta, \, \nu, \, \beta$ -- can be
unambiguously estimated. In this regard, VSP data are different
from the $P$-wave reflection traveltimes that constrain these
five parameters independently only when the tilt $\nu$ exceeds
$30^\circ - 40^\circ$ \protect\cite{gre98}.

We have also assumed that LH is weak. Although this assumption
was not necessary from the standpoint of parameter estimation,
it allowed us to obtain explicit equations describing the
influence of lateral heterogeneity. We showed that these
equations can be inverted and the lateral velocity variation
can be found from traveltimes recorded in a VSP geometry. The
approximate nature of the derived expressions does introduce
some errors in estimated anisotropic parameters; however, the
our numerical examples have demonstrated that those errors are
small compared to the errors produced by applying the conventional
methodology that simply ignores the presence of lateral
heterogeneity.

\section*{Acknowledgments}
We are grateful to members of the A(nisotropy)-Team of the
Center for Wave Phenomena (CWP) at CSM for helpful discussions.
P. Contreras thanks PDVSA-INTEVEP for supporting his visit to
the CWP. The support for this work was provided by the members
of the Consortium Project on Seismic Inverse Methods for
Complex Structures at CWP and by the United States Department
of Energy (award \#DE-FG03-98ER14908).


\end{document}